\documentstyle[twoside]{article}

\catcode`\@=11
\long\def\@makefntext#1{
\protect\noindent \hbox to 3.2pt {\hskip-.9pt  
$^{{\eightrm\@thefnmark}}$\hfil}#1\hfill}		

\def\@makefnmark{\hbox to 0pt{$^{\@thefnmark}$\hss}}	
        
\def\ps@myheadings{\let\@mkboth\@gobbletwo
\def\@oddhead{\hbox{}
\rightmark\hfil\eightrm\thepage}   
\def\@oddfoot{}\def\@evenhead{\eightrm\thepage\hfil
\leftmark\hbox{}}\def\@evenfoot{}
\def\sectionmark##1{}\def\subsectionmark##1{}}

\oddsidemargin=\evensidemargin
\newcounter{sectionc}\newcounter{subsectionc}\newcounter{subsubsectionc}
\renewcommand{\section}[1] {\vspace{12pt}\addtocounter{sectionc}{1} 
\setcounter{subsectionc}{0}\setcounter{subsubsectionc}{0}\noindent 
        {\tenbf\thesectionc. #1}\par\vspace{5pt}}
\renewcommand{\subsection}[1] {\vspace{12pt}\addtocounter{subsectionc}{1} 
        \setcounter{subsubsectionc}{0}\noindent 
        {\bf\thesectionc.\thesubsectionc. {\kern1pt \bfit #1}}\par\vspace{5pt}}
\renewcommand{\subsubsection}[1] {\vspace{12pt}\addtocounter{subsubsectionc}{1}
        \noindent{\tenrm\thesectionc.\thesubsectionc.\thesubsubsectionc.
        {\kern1pt \tenit #1}}\par\vspace{5pt}}
\newcommand{\nonumsection}[1] {\vspace{12pt}\noindent{\tenbf #1}
        \par\vspace{5pt}}

\newcounter{appendixc}
\newcounter{subappendixc}[appendixc]
\newcounter{subsubappendixc}[subappendixc]
\renewcommand{\thesubappendixc}{\Alph{appendixc}.\arabic{subappendixc}}
\renewcommand{\thesubsubappendixc}
	{\Alph{appendixc}.\arabic{subappendixc}.\arabic{subsubappendixc}}

\renewcommand{\appendix}[1] {\vspace{12pt}
        \refstepcounter{appendixc}
        \setcounter{figure}{0}
        \setcounter{table}{0}
        \setcounter{lemma}{0}
        \setcounter{theorem}{0}
        \setcounter{corollary}{0}
        \setcounter{definition}{0}
        \setcounter{equation}{0}
        \renewcommand{\thefigure}{\Alph{appendixc}.\arabic{figure}}
        \renewcommand{\thetable}{\Alph{appendixc}.\arabic{table}}
        \renewcommand{\theappendixc}{\Alph{appendixc}}
        \renewcommand{\thelemma}{\Alph{appendixc}.\arabic{lemma}}
        \renewcommand{\thetheorem}{\Alph{appendixc}.\arabic{theorem}}
        \renewcommand{\thedefinition}{\Alph{appendixc}.\arabic{definition}}
        \renewcommand{\thecorollary}{\Alph{appendixc}.\arabic{corollary}}
        \renewcommand{\theequation}{\Alph{appendixc}.\arabic{equation}}
        \noindent{\tenbf Appendix \theappendixc #1}\par\vspace{5pt}}
\newcommand{\subappendix}[1] {\vspace{12pt}
        \refstepcounter{subappendixc}
        \noindent{\bf Appendix \thesubappendixc. {\kern1pt \bfit #1}}
	\par\vspace{5pt}}
\newcommand{\subsubappendix}[1] {\vspace{12pt}
        \refstepcounter{subsubappendixc}
        \noindent{\rm Appendix \thesubsubappendixc. {\kern1pt \tenit #1}}
	\par\vspace{5pt}}

\newcommand{\textlineskip}{\baselineskip=13pt}
\newcommand{\smalllineskip}{\baselineskip=10pt}

\def\eightcirc{
\begin{picture}(0,0)
\put(4.4,1.8){\circle{6.5}}
\end{picture}}
\def\eightcopyright{\eightcirc\kern2.7pt\hbox{\eightrm c}} 

\newcommand{\copyrightheading}[1]
        {\vspace*{-2.5cm}\smalllineskip{\flushleft
        {\footnotesize International Journal of Modern Physics C, #1}\\
        {\footnotesize $\eightcopyright$\,\,\, World Scientific Publishing
         Company}\\
         }}

\def\abstracts#1#2#3{{
        \centering{\begin{minipage}{4.5in}\baselineskip=10pt\footnotesize
        \parindent=0pt #1\par
        \parindent=15pt #2\par
        \parindent=15pt #3\par
        \end{minipage}}\par}} 

\def\keywords#1{{
       \centering{\begin{minipage}{4.5in}\baselineskip=10pt\footnotesize
       {\footnotesize\it Keywords}\/: #1
        \end{minipage}}\par}}

\renewenvironment{thebibliography}[1]
        {\frenchspacing
	 \ninerm\baselineskip=11pt
         \begin{list}{\arabic{enumi}.}
        {\usecounter{enumi}\setlength{\parsep}{0pt}     
         \setlength{\leftmargin 17pt}{\rightmargin 0pt}   
         \setlength{\itemsep}{0pt} \settowidth
	{\labelwidth}{#1.}\sloppy}}{\end{list}}

\newcounter{itemlistc}
\newcounter{romanlistc}
\newcounter{alphlistc}
\newcounter{arabiclistc}

\newcommand{\fcaption}[1]{
        \refstepcounter{figure}
	\setbox\@tempboxa = \hbox{\footnotesize Fig.~\thefigure. #1}
	\ifdim \wd\@tempboxa > 5in
           {\begin{center}
	\parbox{5in}{\footnotesize\smalllineskip Fig.~\thefigure. #1}
            \end{center}}
        \else
             {\begin{center}
	     {\footnotesize Fig.~\thefigure. #1}
              \end{center}}
        \fi}

\newcommand{\tcaption}[1]{
        \refstepcounter{table}
	\setbox\@tempboxa = \hbox{\footnotesize Table~\thetable. #1}
        \ifdim \wd\@tempboxa > 5in
           {\begin{center}
         \parbox{5in}{\footnotesize\smalllineskip Table~\thetable. #1}
            \end{center}}
        \else
             {\begin{center}
	     {\footnotesize Table~\thetable. #1}
              \end{center}}
        \fi}

\def\@citex[#1]#2{\if@filesw\immediate\write\@auxout
	{\string\citation{#2}}\fi
\def\@citea{}\@cite{\@for\@citeb:=#2\do
	{\@citea\def\@citea{,}\@ifundefined
	{b@\@citeb}{{\bf ?}\@warning
	{Citation `\@citeb' on page \thepage \space undefined}}
	{\csname b@\@citeb\endcsname}}}{#1}}

\newif\if@cghi
\def\cite{\@cghitrue\@ifnextchar [{\@tempswatrue
	\@citex}{\@tempswafalse\@citex[]}}
\def\citelow{\@cghifalse\@ifnextchar [{\@tempswatrue
	\@citex}{\@tempswafalse\@citex[]}}
\def\@cite#1#2{{$\null^{#1}$\if@tempswa\typeout
	{IJCGA warning: optional citation argument 
	ignored: `#2'} \fi}}

\def\pmb#1{\setbox0=\hbox{#1}
        \kern-.025em\copy0\kern-\wd0
        \kern.05em\copy0\kern-\wd0
        \kern-.025em\raise.0433em\box0}

\def\fnt#1#2{\footnotetext{\kern-.3em
        {$^{\mbox{\scriptsize #1}}$}{#2}}}

\def\fpage#1{\begingroup
\voffset=.3in
\thispagestyle{empty}\begin{table}[b]\centerline{\footnotesize #1}
        \end{table}\endgroup}

\def\runninghead#1#2{\pagestyle{myheadings}
\markboth{{\protect\footnotesize\it{\quad #1}}\hfill}
{\hfill{\protect\footnotesize\it{#2\quad}}}}
\font\tenbf=cmbx10
\font\tenit=cmti10 
\font\tenit=cmti10
\font\bfit=cmbxti10 at 10pt

\font\ninerm=cmr9

\font\eightrm=cmr8

\def\lsym{\raise-3pt\hbox{\vbox{\tabskip0pt\offinterlineskip
	\halign{\tabskip0pt plus 1em
	##\tabskip0pt\cr
	$\,\,<\,\,$\cr
	$\,\,\sim\,\,$\cr}}}}
\def\rsym{\raise-3pt\hbox{\vbox{\tabskip0pt\offinterlineskip
     \halign{\tabskip0pt plus 1em
      ##\tabskip0pt\cr
      $\,\,>\,\,$\cr
      $\,\,\sim\,\,$\cr}}}}
\def\qed{\hbox{${\vcenter{\vbox{			
	\hrule height 0.4pt\hbox{\vrule width 0.4pt height 6pt
	\kern5pt\vrule width 0.4pt}\hrule height 0.4pt}}}$}}
\def\theequation{\thesection.\arabic{equation}}		

\begin{document}

\runninghead{R. A. Zara \& R. N. Onody}
{The Optimized Model of Multiple Invasion Percolation}

\normalsize\textlineskip
\thispagestyle{empty}
\setcounter{page}{1}

\copyrightheading{Vol. 0, No. 0 (1998) 000--000}

\vspace*{0.88truein}

\fpage{1}
\centerline{\bf THE OPTIMIZED MODEL OF MULTIPLE }
\vspace*{0.035truein}
\centerline{\bf INVASION PERCOLATION } 
\vspace*{0.37truein}
\centerline{\footnotesize REGINALDO A. ZARA and ROBERTO N. ONODY} 
\vspace*{0.015truein}
\centerline{\footnotesize\it Departamento de F\'{\i}sica e Inform\'{a}tica }
\centerline{\footnotesize\it Instituto de F\'{\i}sica de S\~{a}o Carlos} 
\baselineskip=10pt
\centerline{\footnotesize\it Universidade de S\~{a}o Paulo - Caixa Postal 369}
\centerline{\footnotesize\it 13560-970 - S\~{a}o Carlos, S\~{a}o Paulo, Brasil}
\vspace*{0.15truein}


\vspace*{0.21truein}
\abstracts{ We study the optimized version of the multiple invasion percolation model. 
Some topological aspects as  
the behavior of the acceptance profile, coordination number and  
vertex type abundance were investigated and compared to those of the 
ordinary invasion. Our results indicate that the clusters show a very 
high degree of connectivity, spoiling the usual nodes-links-blobs 
geometrical picture.
}{}{}

\vspace*{10pt}


\vspace*{1pt}\textlineskip	
\section{Introduction}		
\vspace*{-0.5pt}
\noindent
When a nonviscous liquid is injected into a porous medium already
filled with a viscous fluid the system can be found in two different 
regimes: one where the dominant forces are of capillary nature and 
another where the viscous forces are predominant. The theoretical 
description of such system is based on two models: diffusion-limited 
aggregation (DLA) \cite{wit-san} and invasion percolation
\cite{wil-wil}. DLA describes the fast displacement situation 
when the viscous forces are dominant. The invasion percolation model is
applied to a slow fluid flow, where the leading forces are of 
capillary nature. The fluid displacement 
follows minimum resistance paths: the smaller pores are filled
or invaded first. 

The original invasion percolation model have been modified in order to bring 
it as close as possible to real world. 
So, the action of an external gravitational field \cite{wil,bir,mea} and the
flux with a privileged direction \cite{ono1,ono2} were incorporated into the 
model. Recently \cite {ono3} the multiple invasion percolation model was 
proposed. In this model a certain number of lattice sites can be 
simultaneously invaded. It is always important to have in mind
that invasion percolation is a dynamical model, that is, it intends to explain 
not only static properties like the clusters fractal dimensions but also how 
these clusters evolve in time. The lack of a meaning on associating or trying 
to make a correspondence between one step growth interval (of the algorithm) 
and one time interval separating two sucessives pore's invasions (of the real 
world) is what we were concerned by the time we formulated the multiple 
invasion percolation model \cite{ono3}. In our models, one real time 
interval corresponds exactly to one growth step and it is, in this sense, more 
realistic.
When water penetrates layered soil, in some cases, a 
fingering formation is detected. The multiple invasion percolation not only
describes successfully the fingers dynamics but also led us to a more ambitious
program - to obtain the experimental pores size distribution of the soil  
\cite{ono1}. Briefly, the 
multiple invasion percolation model is useful not only to explain experimental
results but also to predict them.

There are two kinds of multiple invasion percolation:
the perimeter and  optimized
models. In the first model the cluster growth is controlled by the flux 
through the perimeter. The optimized model is governed by a scaling relation 
between the mass and the gyration radius of the cluster. Both types of 
invasion were studied in their site 
versions. For the perimeter model, the acceptance profile,
mean coordination number and abundance of vertice type were investigated.
In this paper we discuss the acceptance profile and some topological 
properties of the optimized model clusters.

\setcounter{section}{2}
\setcounter{equation}{0}
\section{The Optimized Model}

We briefly recall the growth mechanism of the optimized model\cite{ono3}. 
It was devised to obey exactly the scaling
 
\begin{equation}
M \sim (R_g)^{D} 		
\end{equation}
or as near it as possible ($R_g$ is the gyration radius and $D$ is
a real positive external parameter that can be tuned). 

Basically we use the following strategy: at each growing step we build a list
containing all the cluster perimeter sites that can be invaded
and we ask for the number of sites that should be invaded in order
that equation (2.1) is verified as closely as possible. When $D \in [1.89,2]$
these proceedings build a fractal object \cite{man} which is extremely 
stabilized (in the sense that in {\em any} stage or size the scaling is 
perfectly obeyed and not only in the asymptotic limit). In this interval the 
parameter $D$ coincides with the fractal dimension of the cluster.  Reference 
\cite{ono3} stress that although 
the ordinary invasion percolation and the optimized model ($D = 1.89$) have 
the same fractal dimension the clusters are not the same. While in the 
ordinary invasion percolation only one site is invaded at each step, in the 
optimized model several 
sites can be invaded at same time. The result is that the optimized model 
generates 
more massive clusters than those of the ordinary invasion percolation.

In the region $ [0,1.89]$ the system is frustrated from below, that is, the 
scaling relation recommends an invasion of less than one site which is 
forbidden by the algorithm 
(at least one site must be invaded). In the region 
$D > 2$, the system is frustrated from above and a very beautiful burst 
phenomenon takes place \cite{ono3}. In both cases, $D$ does not coincide 
with the real fractal dimension.

\section{Numerical Simulations}

In this paper we have studied the optimized model only in the region 
$[1.89,2]$.
To analyze growth mechanisms it is useful to define \cite{wil-wil} the
acceptance profile {\em a(r)} which is the ratio between the number
of random numbers in the interval [ {\em r, r+dr } ] accepted 
into the cluster and the total number of random numbers in that range.
In the limit of an infinite lattice, the acceptance profile of the 
ordinary invasion percolation tends to a step function with the 
discontinuity located at the critical ordinary percolation 
threshold $p_{c} = 0.5928 $ \cite{stau}.
For the optimized model we determined the acceptance profile as a
funtion of $D$. Fig. 1 shows the case $D = 1.89 $ for many
lattice sizes. It does not seems to approach a step function, exhibiting 
a persistent tail near the inflexion point.
This happens because,in order to obey the scaling rule, sites with larger 
random numbers are invaded. As a consequence, we find that the inflexion 
point position is bigger than its value in the ordinary invasion.
The inset of the Fig. 1 shows that the inflexion point position increases 
with $D$. 

\begin{figure}
\vspace{8cm}
\includegraphics{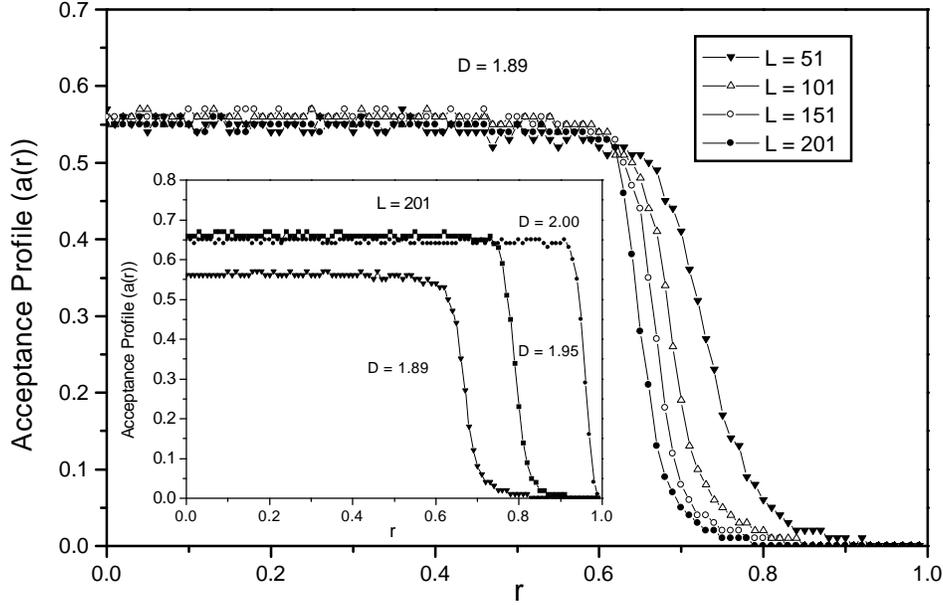}
\caption{ Some acceptance profiles of the optimized model for fixed $D$ and
various lattice sizes. In the inset, we show the behaviour of $a(r)$ for 
$L= 201$ and different values of $D$.} 
\end {figure}

The sites of the multiple invasion percolation cluster can be classified 
according to their number of first nearest neighbors occupied sites. 
Let $N_{k}$ be the number of sites surrounded by $k$ ( $k=1,..,4$ for the 
square lattice) occupied sites. We determined the vertice concentrations of 
kind $k$, i.e., $n_{k}=\frac{N_{k}}{\sum_{k}N_{k}}$, for
sizes  $L=51, 101, 151$  and $201$. The result was extrapolated using the BST
algorithm \cite{bst} and it is presented in the Table 1.
The BST is a useful algorithm to extrapolate physical quantities that converge
obeying a power law $ F(L) = F(\infty) + A L^{-\theta}$. It allows a reliable
determination of critical parameters in the thermodynamic limit and its 
versatility becomes more pronounced if there are only very short sequences
available.   
Our estimated coordination number for the optimized model ($D = 1.89$) 
is $Z=2.71$ in contrast with $Z = 2.51$ for the ordinary invasion 
percolation \cite{ono3}.  In 
the table, we see that, as we increase $D$, the cluster becomes more compact 
favouring vertices of kind $4$ in detriment of those of kind $1$ which are 
practically extinct.

\begin{table}
\begin{center}
\begin{tabular}{|c|c|c|c|c|c|}
\hline
$D$ & $n_{1}$ & $n_{2}$ & $n_{3}$ & $n_{4}$ & $Z$ \\ \hline
1.89 & 0.10 & 0.30 & 0.40 & 0.20 & 2.71 \\ \hline
1.92 & 0.07 & 0.26 & 0.41 & 0.26 & 2.78 \\ \hline
1.95 & 0.04 & 0.17 & 0.41 & 0.38 & 3.14 \\ \hline
1.98 & 0.02 & 0.08 & 0.32 & 0.58 & 3.48 \\ \hline
2.00 & 0.01 & 0.01 & 0.10 & 0.88 & 3.88 \\ \hline
\end{tabular}
\caption { The vertice concentrations $n_{k}=\frac{N_{k}}{\sum_{k}N_{k}}$ and the coordination 
number $Z$ of the optimized model for many values of $ D $. The data were
extrapolated using the BST algorithm.}
\end{center}
\end{table}

The topological properties of the percolating cluster 
have been investigated for more than two decades. It is well known that the 
backbone is an 
important structure of the cluster. It is formed by the union of 
all self-avoiding walks connecting two points $P_{1}$ and $P_{2}$ of the 
lattice. This means that
if we pass a current between these points, the backbone is 
identified as the set of sites that carries current. The backbone study
has possible applications on the condutivity of random 
systems \cite{gen} and on fluids flowing in porous media \cite{sta-co}. 
The backbone of the critical percolation cluster is non-Euclidean with 
fractal dimension $ 1.647 \pm 0.004 $ \cite{gras}. 

Many pictures of the percolating cluster backbone were proposed 
\cite{gen,skal,gef}. 
The most promising model was introduced by Stanley \cite{blob1} and is know 
as the {\em nodes-links-blobs model}. In this model the backbone consists of 
a network of nodes connected by one-dimensional links which are often 
separated by multiconnected pieces or blobs of all length scales \cite{blob2}.
Thus, the backbone may be viewed as a topologically linear string of blobs of
all possible sizes.
In general, there are sites that, when removed, split the spanning 
cluster into pieces. These sites are termed red sites. In the 
nodes-links-blobs model the red sites are also called blobs of size one 
\cite{blob2}. At criticality, the number of red sites scales as a power 
law with the exponent calculated exactly by Coniglio \cite{con} 
$\frac{1}{\nu} = 0.75$.

We have already studied the backbone and elastic backbone structures of the 
multiple invasion percolation model \cite{zara}. To determine these structure
we employed the burning algorithm \cite{herr}. We chose this algorithm 
because, beyond the backbone and the elastic backbone, it also permits the 
determination of the red sites number, the minimum path and the loops number. 

\begin{figure}
\vspace{7.0cm}
\includegraphics{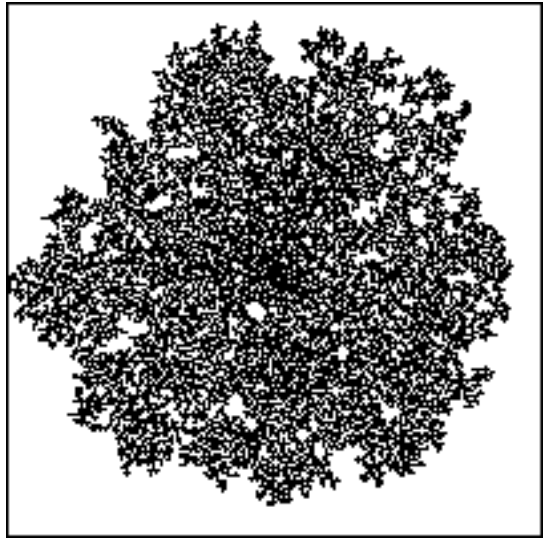}
\caption{ A typical cluster of the optimized model for $D = 1.89$ and $L=201$.
The cluster has 14650 sites.} 
\end {figure} 

For the optimized model, the scaling (equation (2.1)) is also perfectly 
obeyed by the backbone. At $D=1.89$ we got the fractal dimension 
$D_F=1.74 \pm 0.01$. This means that although the optimized
model at $D=1.89$ and the ordinary invasion have the {\em same}  
fractal dimension, they are intrinsicaly different since their backbones 
are not the same. The number of red sites $N_{r}$ is very small and random. 
It does not seem to obey any power law.
The red sites number is so small that the probability of disconnecting
the cluster by removing any site randomly is pratically zero. The optimized 
algorithm destroys the red sites, increasing the cluster connectivity. The
disappearence of the red sites indicates that the nodes-links-blobs model 
may not be useful to describe the backbones of the optimized model clusters. 
Indeed, the cluster of the optimized model has a very different form from 
that of the ordinary 
invasion percolation as can be seen in the Fig. 2. 

\section{Conclusions}

We studied the optimized model of multiple invasion percolation
by  comparing its topological properties with those of the ordinary invasion 
percolation. The high connectivity of the clusters produced by the 
optimized model together with the almost disappearance
of the red sites, led us to conclude that the nodes-links-blobs model is 
not well suited to describe the backbones of the optimized model.     

\section{Acknowledgements}

We acknowledge CNPq (Conselho Nacional de Desenvolvimento
Cient\'{\i}fico e Tecnol\'ogico) and FAPESP ( Funda\c c\~ao de Amparo
a Pesquisa do Estado de S\~ao Paulo ) for the financial support.

\nonumsection{References}

\end{document}